\documentclass[twocolumn,aps,showpacs,preprintnumbers,amsmath,amssymb,superscriptaddress]{revtex4-1}
\usepackage{graphicx}
\usepackage{dcolumn}
\usepackage{bm}
\usepackage{amsmath}
\let\oldAA\AA
\renewcommand{\AA}{\text{\normalfont\oldAA}}
\usepackage{enumitem}
\usepackage{float}

\usepackage{amsmath}
\usepackage{subfigure}
\usepackage{hyperref}
\hypersetup{
    colorlinks,
    citecolor=blue,
    filecolor=black,
    linkcolor=blue,
    urlcolor=blue
}
\usepackage{epstopdf}
 \usepackage{relsize}

\begin{document}


\title{Performance analysis of nanostructured Peltier coolers}

\author{Aniket Singha}
 \email{aniket@iitbhilai.ac.in}
\affiliation{%
Department of Electrical Engineering,\\
Indian Institute of Technology Bhilai, Datrenga, Raipur-492015, India\\
}%
\affiliation{%
Department of Electrical Engineering,\\
Indian Institute of Technology Bombay, Powai, Mumbai-400076, India\\
}%
\author{Bhaskaran Muralidharan}%
 \email{bm@ee.iitb.ac.in}
\affiliation{%
Department of Electrical Engineering,\\
Indian Institute of Technology Bombay, Powai, Mumbai-400076, India\\
}%



\date{\today}

\begin{abstract}
Employing non-equilibrium quantum transport models, we investigate the details and operating conditions of nano-structured Peltier coolers embedded with an energy filtering barrier. Our investigations point out non-trivial aspects of Peltier cooling which include an inevitable trade-off between the cooling power and the coefficient of performance, the coefficient of performance being high at a low voltage bias and subsequently deteriorating with increasing voltage bias. We point out that there is an optimum energy barrier height for nanowire Peltier coolers at which the cooling performance is optimized. However, for bulk Peltier coolers, the cooling performance is enhanced with the height of the energy filtering barrier. Exploring further, we point out that a degradation in cooling performance with respect to bulk is inevitable as a single moded nanowire transitions to a multi-moded one.  The results discussed here can provide theoretical insights for optimal design of nano Peltier coolers. 
\end{abstract}

\maketitle


\section{Introduction}
In the current nanotechnology era, the rise in operating temperatures of nanodevices as a result of increasing dissipated heat density has revived an interest in effective heat management and Peltier coolering. With the recent discovery of thermoelectric materials with high figures of merit \cite{highfom1, highfom2, thermoinnano,aniket,anifil}, there has been a lot of theoretical and experimental effort in an attempt to meet the demand for high performance Peltier coolers \cite{snyder_thompson,cooling_ref1,cooling_ref2,cooling_ref3,cooling_ref4,cooling_ref5,cooling_ref6,cooling_ref7,cooling_ref8,cooling_ref9,cooling_ref10,cooling_ref11}. Peltier cooling is facilitated by an energy selective disturbance in quasi-equilibrium among the electronic population via energy filtering. Such a disturabance in quasi-equilibrium, in conjugation with inelastic processes, initiates heat absorption from the lattice \cite{snyder_thompson,cooling_ref2,cooling_ref5,cooling_ref6,cooling_ref9,cooling_ref10,cooling_ref11,whitney2,whitney}.   Despite attempts towards the theoretical and experimental realization of high performance Peltier coolers \cite{snyder_thompson,cooling_ref1,cooling_ref2,cooling_ref3,cooling_ref4,cooling_ref5,cooling_ref6,cooling_ref7,cooling_ref8,cooling_ref9,cooling_ref10,cooling_ref11}, an overall analysis of the functionality and optimum operating conditions of nano Peltier coolers is missing in the current literature. In this paper, we hence study the performance of various nanoscale Peltier coolers in order to analyze the optimum operating conditions. \\
\indent There are two major pathways to facilitate an overall performance improvement in Peltier coolers: (i) decreasing the lattice heat conductivity (ii) enhancing the cooling power via suitable energy filtering techniques. In the last few decades, approaches towards nano-structuring, hetero-structuring and density of states engineering have so far proven successful in the suppression of phonon mediated lattice thermal conductivity via scattering and confinement of long wavelength phonons \cite{phonon1,phonon2,phonon3,phonon4,phonon5,phonon6,phonon7,superlattice1,superlattice2,nanoflake_heat,nanowire_heat1,nanowire_heat2}. Hence, in this paper, we   explore the other aspect, that is, enhancing the cooling power in nanostructures. In particular, we explore  Peltier cooling in nanostructures embedded with an  energy filtering barrier. \\
\indent In the aspect of Peltier cooling, we believe that a few points deserve special attention. First of all, the few recent theoretical works \cite{ieeecool,snyder_thompson,cooling_ref1,cooling_ref2,cooling_ref6}  in this aspect are based on a linear response analysis. Linear response analysis masks the essential combination of transport physics and scattering events that jointly determine the net cooling power as well as the coefficient of performance ($COP$). In addition, the linear response limit is broken in the regions which strongly deviate from quasi equilibrium, namely in the vicinity of the energy filtering barrier. Secondly, a generalized picture of the physics of cooling performance is unclear from the available literature. With respect to the first point, our analysis of cooling performance is based on the non-equilibrium Green's function formalism which accounts for the non-equilibrium nature of transport to directly evaluate the charge and heat currents.  \\
\indent This paper is organized as follows. First, we briefly elaborate the underlying physics of Peltier cooling in Sec.~\ref{explain} following which we briefly describe the  transport formulation in Sec.~\ref{transport}. We next elaborate our study on Peltier cooling in nanostructures in Sec.~\ref{results} where we mainly analyze the  \emph{ cooling power} and  the \emph{ COP} at a chosen cooling power. We show that the cooling power increases while the $COP$ decreases as one increases the applied bias voltage depicting  a trade-off between the the two. Exploring further, we demonstrate that a deterioration in cooling performance with respect to bulk is unavoidable as a single-moded nanowire transitions to a multi-moded regime.  We end the paper with a general conclusion in Sec.~\ref{conclude}. The transport formalism used for the simulations is detailed in the Appendix at the end of this paper. 

\begin{figure}
\includegraphics[scale=1.2]{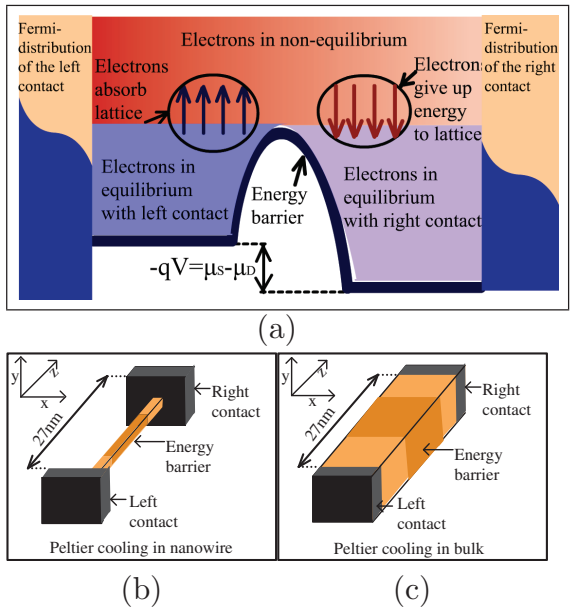}
\caption{Schematic diagram showing the phenomenon of Peltier cooling. The electrons at the source side of the barrier tend to equilibriate by absorbing heat from the lattice while those at the drain side of the barrier tend to equilibriate by releasing heat to the lattice.}
\label{fig:cooling_schematic}
\end{figure}

\section{Peltier cooling in semiconductor heterostructures}\label{explain}
Thermoelectric or Peltier cooling in semiconductor heterostructures is facilitated by giving rise to an energy selective lack of quasi-equilibrium among the electronic population  via energy filtering. In the classical limit, when a potential is applied across the barrier, electrons above the energy barrier height $E_b$ tend to migrate from the source contact to the drain contact giving rise to  a  local non-equilibrium among the high energy electronic population. The electronic population below the energy barrier height $E_b$, however, remains in quasi-equilibrium with the respective contacts. This energy-selective lack of quasi-equilibrium initiates a heat absorption process from the lattice via inelastic scattering. In the case of a Peltier refrigerator  embedded between  macroscopic contacts, two equivalent phenomena give rise to cooling and heating at the two interfaces of the energy filtering barrier as demonstrated in Fig. \ref{fig:cooling_schematic}~(a). (i) The high energy electrons at the source side of the barrier interface  are driven out of equilibrium initiating heat absorption in that region. (ii) The high energy electrons migrating towards the drain side are out of equilibrium with the drain quasi-Fermi potential due to the external voltage bias ($\mu_D=\mu_S-V$). Hence, the  electrons energetically relax giving up heat energy to the lattice. The efficacy of a Peltier refrigerator is measured by the $COP$ ($\zeta$) defined as: 
\begin{equation}
\zeta=J_C/P,
\label{eq:COP}
\end{equation} 
where, $J_C$ is the rate of electronic heat extraction from the source side of the barrier interface, or equivalently, the rate of Peltier cooling and  $P$ is the power consumed from the external source given by,
 \[
 P=V\times I,
 \]
 $V$ being the applied bias and $I$ being the current  flowing through the Peltier refrigerator. The rate of heat extracted from the source side of the barrier interface ($J_C$) as well as the the $COP$ ($\zeta$),  depend on the position of the equilibrium Fermi potential ($\mu_0$) or equivalently, the electrochemical potential with respect to the height of the energy filtering barrier.
\section{Transport formulation and model }\label{transport}
\subsection{Transport formulation}
\indent To perform the  calculations, we employ the NEGF transport formalism with inelastic scattering incorporated via the self-consistent Born approximation \cite{dattabook,Datta_Green,LNE}  (details given in the Appendix). The single particle Green's function $G(\overrightarrow{k_{m}},E)$, for each transverse sub-band $m$ \cite{dattabook}, can be   calculated from the device Hamiltonian  $[H]$:
\begin{gather}\label{eq:negf_main}   
G(\overrightarrow{k_{m}},E)=[EI-H-U-E_m-\Sigma(\overrightarrow{k_{m}},E)]^{-1} ,\nonumber \\
\Sigma(\overrightarrow{k_{m}},E)=\Sigma_L(\overrightarrow{k_{m}},E)+\Sigma_R(\overrightarrow{k_{m}},E)+\Sigma_S(\overrightarrow{k_{m}},E), 
\end{gather}
where $[H]$ is the device  Hamiltonian matrix constructed with effective mass approach \cite{dattabook,Datta_Green} and $I$ is the identity matrix of identical order as the Hamiltonian. The spatial profile of  the conduction band minimum is described by the matrix $U$, while  $E$ is the free variable representing the energy of electronic wavefunction.  The sub-band energy  of the $m^{th}$ sub-band is calculated assuming parabolic $E-k$ dispersion relation: 
\[
E_m=\frac{\hslash^2k_m^2}{2m_t}. 
\]
The wavevector of the electron in the transverse direction for the $m^{th}$ sub-band is denoted by $\overrightarrow{k_{m}}$. The total scattering self-energy matrix $[\Sigma(\overrightarrow{k_{m}},E)]$ incorporates the effect of  scattering of the electronic wavefunctions from the contacts into the active device region, which is represented  by $\Sigma_L(\overrightarrow{k_{m}},E)+\Sigma_R(\overrightarrow{k_{m}},E)$ as well as the scattering of electronic wavefunctions inside the device due to 
inelastic processes, which is  denoted by  $\Sigma_S(\overrightarrow{k_{m}},E) $ (detailed in the Appendix). The   scattering functions are calculated self-consistently with the transport calculations, (detailed in the Appendix), with the electron and the hole density operators $G^n(\overrightarrow{k_{m}},E)$, $G^p(\overrightarrow{k_{m}},E)$ given by
\begin{eqnarray}
G^n(\overrightarrow{k_{m}},E)=G(\overrightarrow{k_{m}},E)\Sigma^{in}(\overrightarrow{k_{m}},E)G^{\dagger}(\overrightarrow{k_{m}},E), \nonumber \\
G^p(\overrightarrow{k_{m}},E)=G(\overrightarrow{k_{m}},E)\Sigma^{out}(\overrightarrow{k_{m}},E)G^{\dagger}(\overrightarrow{k_{m}},E). 
\end{eqnarray}
On the  convergence of the self-consistent calculations, the charge and heat currents propagating from the $j^{th}$ lattice point to the $(j+1)^{th}$ lattice point are computed 
as:
\begin{multline}
 I^{j\rightarrow j+1}_C  =\underset{k_m}{\sum}i\frac{e}{\pi \hslash} \int[
G^n_{j+1,j}(\overrightarrow{k_{m}},E)H_{j,j+1}(E)    
\\ -H_{j+1,j}(E)G^n_{j,j+1}(\overrightarrow{k_{m}},E) ]dE,  \nonumber 
\end{multline}
\begin{multline}
 I_Q^{j\rightarrow j+1}  =\underset{k_m}{\sum}\frac{i}{\pi \hslash} \times  \int E[
G^n_{j+1,j}(\overrightarrow{k_{m}},E)  H_{j,j+1}(E) 
\\ -H_{j+1,j}(E)G^n_{j,j+1}(\overrightarrow{k_{m}},E) ]dE,  
\label{eq:heatcurrentnegf}
\end{multline}
 where $M_{i,j}$, in the above set of equations, denotes a generic matrix element of the  operator $M$ between two lattice points $i$ and $j$. In the nearest neighbour tight-binding approximation used here, we only consider the next nearest neighbor such that $j=i \pm 1$. The cooling power per unit volume ($\frac{1}{A} \frac{dJ_C}{dz}$) at the $j^{th}$ point  along the transport direction is then calculated from the equation:
 \begin{equation}
 \frac{1}{A}\frac{dJ_C}{dz}\Bigg|_j=\frac{1}{A}\frac{dI_Q}{dz}\Bigg|_j=\frac{1}{A}\frac{I_Q^{j\rightarrow j+1}-I_Q^{j-1\rightarrow j}}{a},
\end{equation} 
$a$ being the lattice constant used for simulation and $A$ is the cross-sectional area of the Peltier refrigerator. The total cooling power per unit area at the source side of the barrier interface is the given by:
\begin{equation}
\frac{J_C}{A}=\frac{1}{A} \int \frac{dJ_C}{dz} \theta \left(\frac{dJ_C}{dz}\right)dz,
\end{equation}
where $\theta(\iota)$ is the unit step function with argument $\iota$. 
\subsection{Model}
We  perform a detailed analysis of Peltier cooling in nano-wires and bulk. The device structures considered here include nanowires whose transverse extent include  only one sub-band and bulk whose transverse extent is infinite (schematic shown in Fig.~\ref{fig:cooling_schematic}~(b) and (c) respectively). We analyze the cooling power vs. $COP$ for a range of values of the reduced Fermi energy given by:
\begin{equation}
\eta_f=\frac{E_c+E_b-\mu}{k_BT},
\label{eq:reduced_mu}
\end{equation}
where $\mu$ is the equilibrium Fermi potential and $E_b$ is the height of the energy barrier. For the purpose of simulation, we use the parameters of $\Delta_2$ valley of lightly doped silicon \cite{book1}, the longitudinal effective mass being $m_l=m_e$ and the transverse effective mass being $m_t=0.2m_e$ ($m_e$ being the free electron mass). The inelastic processes considered here are assumed to be local with an energy exchange given by  $\hslash \omega=30meV$. The temperature of the entire device is considered to be $T=300K$. Under normal conditions, the device region at the source side of the barrier interface would be cooled while the same in close proximity to the drain side of the barrier interface would be heated. However, we assume that the difference between the maximum and the minimum temperature in the Peltier refrigerator is small compared to the average  temperature and hence an assumption of constant temperature throughout the entire device  is justified. For simplicity, the contacts are assumed to be reflection-less macroscopic bodies with electronic distribution in equilibrium at temperature $T$. Their respective quasi-Fermi potential, labeled as $\mu_S$ and $\mu_D$ respectively, are assumed to be $\mu_{S(D)}=\mu_0 \pm V/2$, where $V$ is the externally applied bias voltage. For the purpose of simulation, the devices are assumed to be embedded with a Gaussian potential barrier of the form:
\[
U=E_bexp\left[-\frac{(z-z_0)^2}{2\sigma _w^2}\right],
\]
where $E_b$ and $\sigma_w$ define the  energy filtering barrier height and width respectively and $z_0=L/2$ is the mid-point of the device, $L$ being the total length of the refrigerator in between the contacts. \\
\begin{figure}
\centering
{\includegraphics[scale=.33]{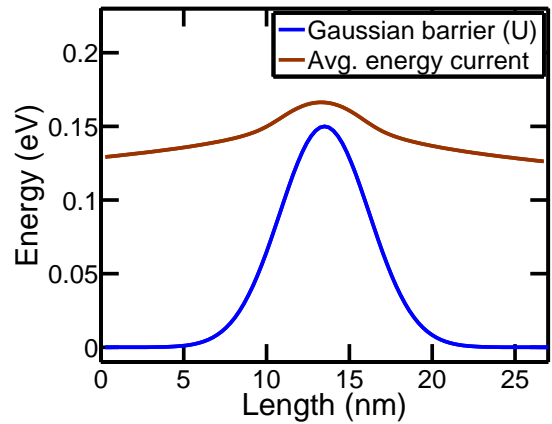}}
\caption{Plot of the embedded energy barrier (blue)   and average energy current (brown) flowing through the nanowire (cross-section=$2.7nm \times 2.7nm$, length=$27nm$) Peltier refrigerator at $V=5~meV$ and $\eta_f=2$.  The embedded energy barrier is assumed to be Gaussian   with height $E_b=150~meV$ and $\sigma_w=2.7~nm$. The average energy current is high near the barrier interface due to absorption of lattice heat. }
\label{fig:energy_current}
\end{figure}
\begin{figure}
\centering
\subfigure[]{\includegraphics[scale=.3]{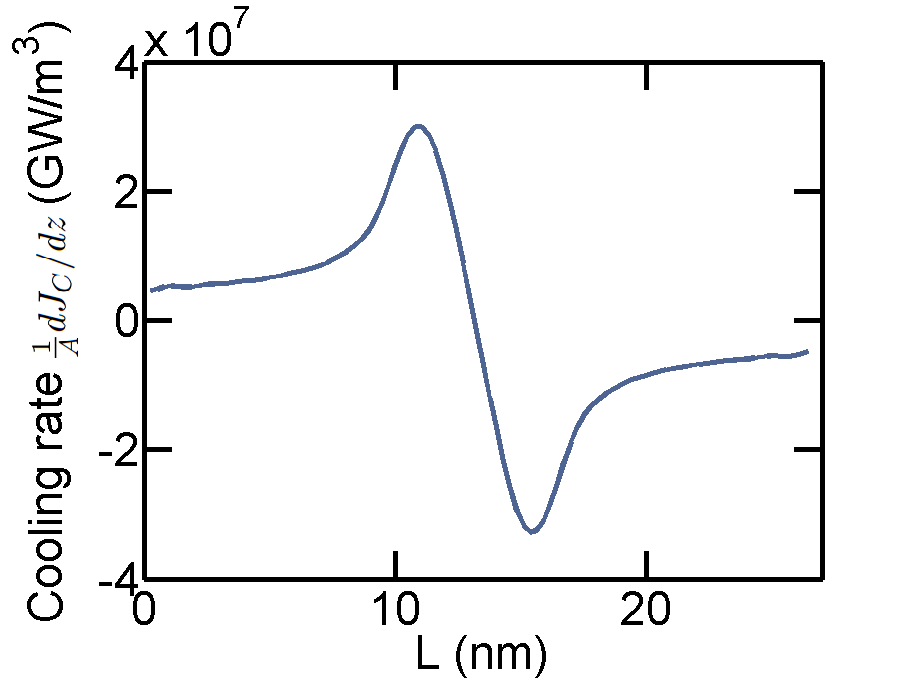}
}
\subfigure[]{\includegraphics[scale=.3]{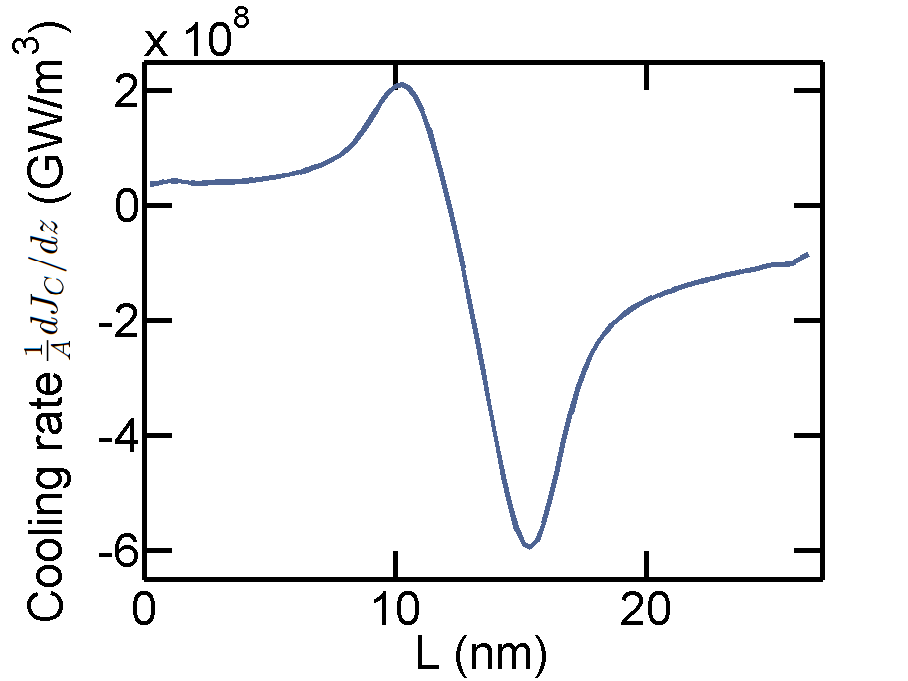}
}
\caption{Spatially resolved cooling profile for a square nanowire of length $27nm$ and width $2.7nm$ (a) when the applied voltage is  much lower compared to $k_BT$ ($V=5meV$)  (b) at the maximum achievable cooling power. The spatial profile of the embedded energy barrier is assumed to be Gaussian ($\sigma_w=2.7nm,~E_b=150meV$).}
\label{fig:cooling_comparable}
\end{figure}
\begin{figure*}[!htb]
\hspace{-1.5cm}\includegraphics[scale=1.6]{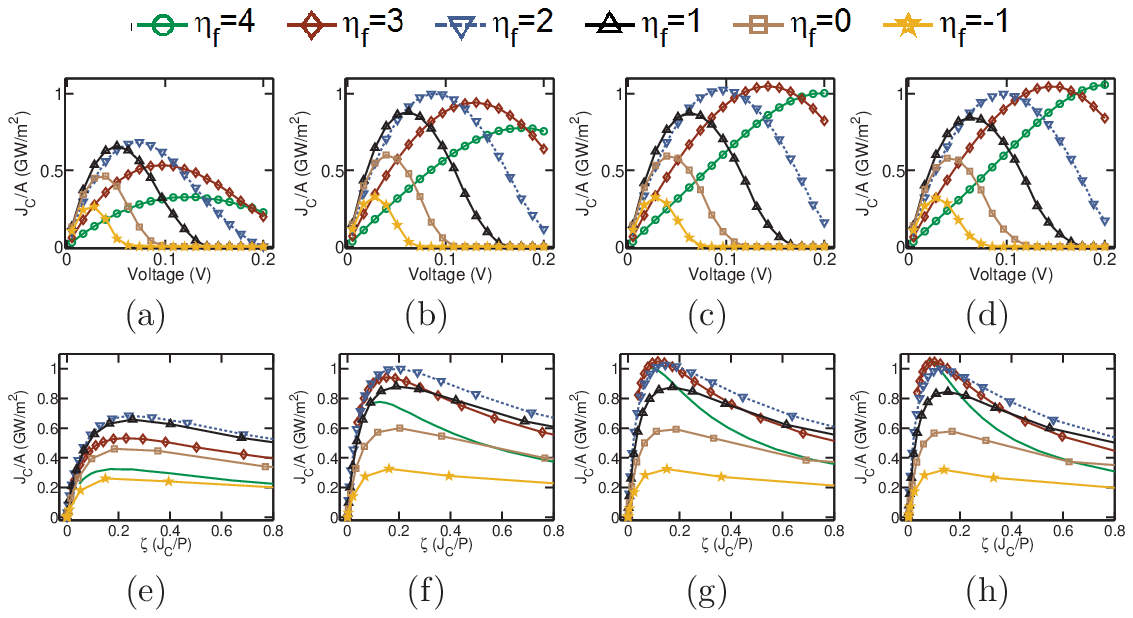}
\caption{Cooling performance analysis for nanowire Peltier coolers at various values of the reduced Fermi energy $\eta_f$.  (a-d) Cooling power at various voltage biases for (a) $E_b=100meV$, (b) $E_b=150meV$, (c) $E_b=200meV$, (d) $E_b=250meV$. (e-h): Analysis of cooling power vs. the $COP$ for (e) $E_b=100meV$, (f) $E_b=150meV$, (g) $E_b=200meV$, (h) $E_b=250meV$.   Simulations for a square $2.82~nm$ nanowire of length $27nm$ embedded with a Gaussian nano-barrier of width $\sigma _w=2.7~nm$. The cooling power vs. voltage curves are plotted for each value of the reduced Fermi energy defined as: $\eta_f=\frac{E_c+E_b-\mu}{k_BT}$.}
\label{fig:nanowire}
\end{figure*}
\begin{figure}[!htb]
\hspace{-1cm}\includegraphics[scale=1]{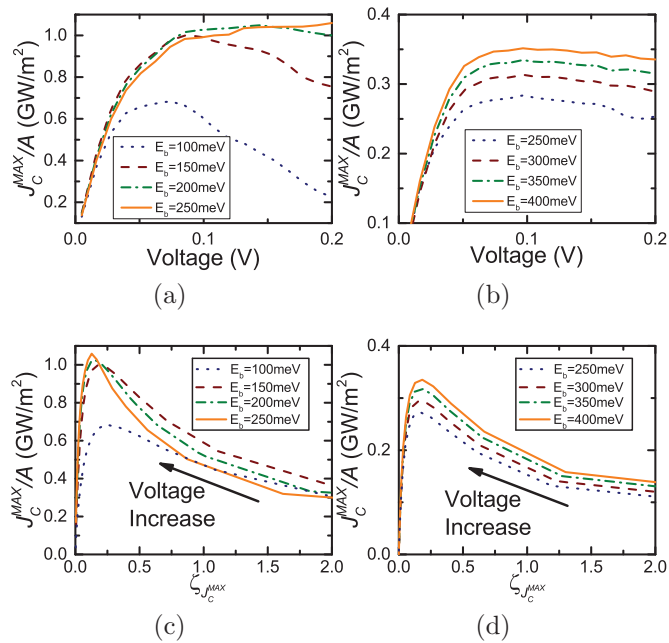}
\caption{Maximum cooling power at a given voltage for (a) nanowire Peltier refrigerator (b) bulk Peltier refrigerator. Operating line characteristics (maximum $COP$ for a given cooling power ) for (c) nanowire Peltier coolers (d) bulk Peltier coolers. The coolers are assumed to be $27nm$ in length with an embedded Gaussian energy barrier ($\sigma_w=2.7nm$)}
\label{fig:op_line}
\end{figure}
\begin{figure}[!htb]
 \includegraphics[scale=1.2]{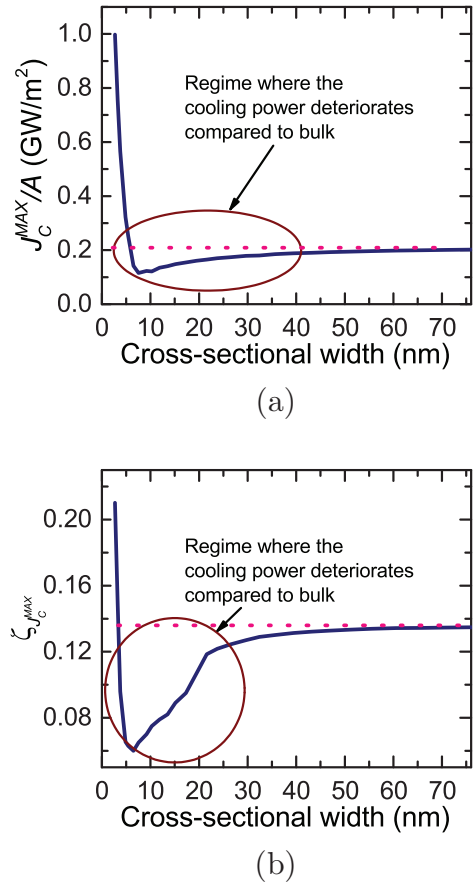}
\caption{The change in maximum cooling power density ($J_C^{MAX}$) and $COP$ at the maximum cooling power density as a single-moded nanowire makes transition towards the bulk regime.}
\label{fig:transit}
\end{figure}
\begin{figure}[!htb]
\includegraphics[scale=0.25]{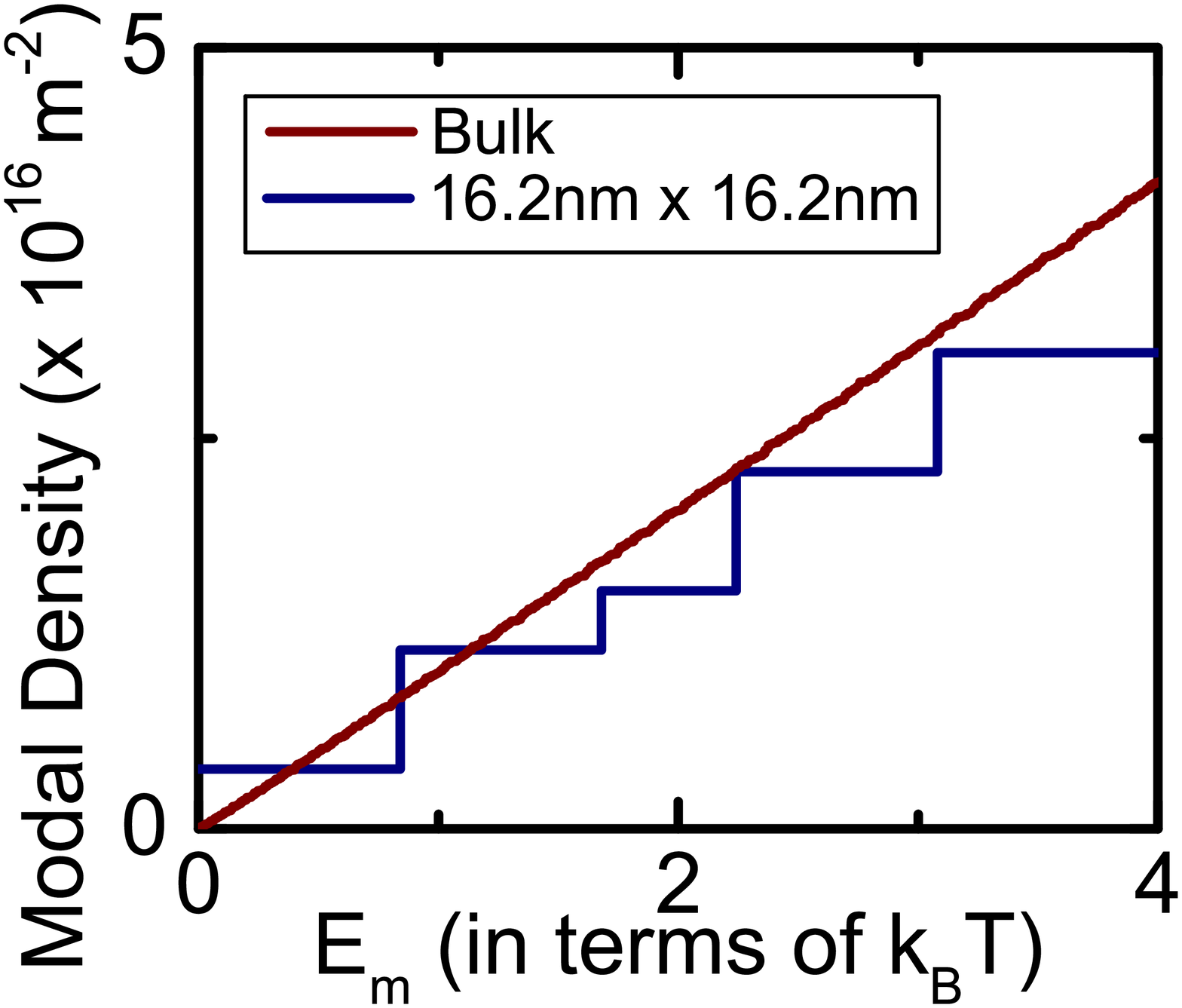}
\caption{The density of modes per unit area of a multimoded nanowire ($16.2nm \times 16.2nm$) and bulk Peltier refrigerator. The density of modes per unit area in the multi-moded regime is less compared to the bulk regime resulting in a deterioration of the maximum cooling power.}
\label{fig:modal}
\end{figure}
\indent Although a more rigorous method would be to solve the potential profile along the device using information on position dependent doping concentration and hetero-junction band-offsets, we believe that our model captures the essential physics and hence, the trends noted in the simulation results  would not deviate drastically with the shape of energy filtering barrier used for simulation. This is because Peltier cooling is dependent on the efficacy of electronic  filtering as well as the rate inelastic scattering and not on the absolute nature of the energy filtering barrier being used.  A list of parameters used for the simulation of the NEGF equations are given in Tab. I. \\
\begin{table}[!htb]
\caption{Parameters used for simulation in this chapter.}
\begin{center}
    \begin{tabular}{ | p{4cm} | p{4cm} |}
    \hline
    \textbf{Parameters} & \textbf{Values}  \\ \hline \hline
    $T~~(k_BT)$ & $300K~~(25.85~meV)$ \\ \hline
    Length of device  & $27nm$ \\ \hline
    $m_l$ & $m_e$  \\ \hline
    $m_t$ & $0.2m_e$ \\ \hline
    $D_O$ (Appendix) & $0.1F~eV^2$  \\ \hline
    $a$ (lattice constant)& $2.7 \AA$ \\ \hline
    $E_c$ (conduction band-edge) & $0eV$ \\ \hline
    $\hslash \omega$ & $30meV$ \\ \hline
    \end{tabular}
       \vspace{1ex}
       
{\scriptsize \emph{\textbf{Note:}} $F=\frac{1}{N_xN_y}$, where $N_x$ and $N_y$ are the number of lattice points in the $x$ and $y$ directions. $m_e$ is the free electron mass and $D_O$ is related to the acoustic deformation potential (See Appendix). }
\end{center}
\end{table}
\section{Results}\label{results}
\indent \textbf{Peltier cooling with inelastic scattering in nanowires:} We first explore Peltier cooling in a single-moded nanowire embedded with an energy filtering barrier. We plot in Fig.~\ref{fig:energy_current} the spatially resolved average energy of the electronic current flowing through the nanowire Peltier refrigerator at a low voltage bias. The average energy is high near the energy barrier interface due to the absorption of lattice heat energy. The  spatial cooling profile of a nanowire Peltier refrigerator is shown in Fig.~\ref{fig:cooling_comparable}.  Particularly,  Fig.~\ref{fig:cooling_comparable}~(a) and (b) demonstrate  the spatial cooling profile ($\frac{1}{A} \frac{dJ_C}{dz}$) when the applied voltage biases are low and high respectively compared to $k_BT$. \\
\indent When the applied voltage $V<<k_BT$, the amount of heat extracted from the source side is almost identical to the amount of heat dissipated at the drain side of the barrier interface (Fig.~\ref{fig:cooling_comparable}~a) resulting a high $COP$. The operating point in such a case is near the reversible limit.  However, the net cooling power under such conditions is low. On the other hand, when the applied bias is high compared to $k_BT$, the heat extracted from the source side of the barrier interface is much less compared to the heat dissipated at the drain side of the barrier interface (Fig.~\ref{fig:cooling_comparable}~b) resulting in a strong deviation from the reversible regime. The $COP$ in such a case is low. Fig. \ref{fig:nanowire} demonstrates the cooling characteristics of a nanowire Peltier cooler. \\
\indent In particular, Figs.~\ref{fig:nanowire}~(a-d) depict the cooling power vs. applied bias characteristics while Figs.~\ref{fig:nanowire}~(e-h) depict the cooling power vs. the $COP$ characteristics for the nanowire cooler for various heights of the energy filtering barrier. We note that the maximum cooling power increases with an increase in the height of the energy barrier upto a saturation point. Such a saturation occurs approximately beyond $E_b=150meV$.  The optimized  $COP$ at a given cooling power, on the other hand,  is achieved when the height of the embedded energy filtering barrier is approximately $150meV$.  The optimum position of the Fermi energy in such a case is given  by $\eta_f=2$. We hence conclude that there is a particular height of the energy filtering barrier at which the performance of the nanowire Peltier cooler is optimized. \\
\indent Two competing phenomena can be responsible for such a behaviour as the energy barrier height is increased: (a) an increase in the charge current due to a decrease in the electronic scattering rate as a result of direct tunneling, and (b) a decrease in the rate of heat absorption per electron from the lattice due to the decrease in the scattering rate.  These two competing phenomena result in a saturation of the cooling power beyond $E_b=150meV$. The slight decrease in the $COP$ at a given cooling power with the increase in barrier height beyond $E_b=150meV$ occurs due to a saturation in the cooling power despite an increase in the electronic current. We also note from Figs.~\ref{fig:nanowire}~(e-h) that the $COP$ decreases with an increase in cooling power indicating a trade-off between the two. The deterioration in cooling power with the increase in the applied potential bias beyond a certain limit occurs  as a result of lowering of the energy filtering barrier due to the external bias voltage. Such a lowering of the potential barrier causes an increase in the direct electronic tunneling rate.\\
\indent \textbf{Operating lines:} In the context of Peltier coolers, we define operating line as the locus of points in the  $J_C-\zeta$ space where the maximum $\zeta$ is obtained for a given cooling power. The operating line is important for practical applications where the design or operating considerations mainly aim to maximize $\zeta$ for a given value of $J_C$. We plot in Figs.~\ref{fig:op_line}~(a) and (b) the maximum cooling power of nanowire and bulk Peltier coolers at a given voltage bias while  Figs.~\ref{fig:op_line}~(c) and (d)  demonstrate the operating lines of a nanowire and bulk Peltier cooler respectively for several heights of the energy filtering barrier. As stated previously, we note that the cooling power for nanowire Peltier coolers (Fig.~\ref{fig:op_line}~a) practically saturates beyond $E_b=150~meV$. The COP $\zeta$ at a given cooling power along the operating line of the nanowire rcooler (Fig.~\ref{fig:op_line}~c), however, is optimized for $E_b=150~meV$. However, we note that the cooling power as well as $\zeta$ in bulk Peltier coolers increase with the increase in energy barrier heights (Fig.~\ref{fig:op_line}~b and d). Such a trend occurs due to a monotonic increase in the density of states as well as inelastic scattering rates (assuming a parabolic dispersion relationship) with energy in the case of bulk Peltier coolers. \\
\indent \textbf{Nanowire to bulk transition in Peltier coolers:} The cooling performance in the transition regime between single-moded nanowire and bulk Peltier coolers is of particular interest. Two quantities which may be used to gauge the performance of Peltier coolers are $(i)$ the maximum cooling power density \Big($\frac{J_C^{MAX}}{A}$\Big) and $(ii)$ the COP $\zeta$ at the maximum cooling power ($\zeta _{J_C^{MAX}}$). We plot in Figs.~\ref{fig:transit}~(a) and (b), the maximum cooling power and the $\zeta$ at the maximum cooling power as a single-moded nanowire gradually transitions to the bulk regime. It is evident from Fig.~\ref{fig:transit}~(a) that a single-moded nanowire  provides an enhanced cooling performance compared to bulk  due to greater conductance per unit area as well as efficient energy filtering due to the abrupt feature in the density of states (the Van Hove singularity).  The variation in cooling performance as a single-moded nanowire transitions to the bulk regime is, however, of particular interest.  The maximum cooling power density ($\frac{J_C^{MAX}}{A}$) as well as the $COP$ at the maximum power  ($\zeta _{J_C^{MAX}}$) of a single-moded nanowire Peltier cooler, demonstrated in Figs.~\ref{fig:transit}~(a) and (b) respectively, deteriorate compared to bulk  as it transitions to the multi-moded regime and subsequently increases toward the bulk values as the multimoded nanowire gradually becomes equivalent to the bulk regime. For large cross-section, the separation between consecutive sub-bands in a nanowire becomes much less than $k_BT$ and the nanowire begins to exhibit bulk properties.\\
\indent Such a behaviour in the maximum power density can be well explained from the modal density profile of a multi-moded nanowire compared to bulk, as shown in Fig.~\ref{fig:modal}. The enhanced cooling power in bulk coolers compared to multi-moded nanowire coolers is a result of higher density of modes in bulk. The degradation in $\zeta _{J_C^{MAX}}$  in the multi-moded regime is not intuitive from a similar argument since $\zeta$ is the ratio between two quantities that are themselves dependent on the modal density profile. However, we noted that the maximum cooling power in multimoded nanowires occurs at a higher bias voltage which, we speculate, leads to a degradation in the $COP$.
\section{Conclusion}\label{conclude}
In this paper, we have analyzed the  cooling performance in nanowire and bulk Peltier coolers. The two parameters we have focused on include the cooling power ($J_C$) and the $COP$  ($\zeta$). We have uncovered some crucial aspects in Peltier coolers which include: (i) there is a trade-off between the cooling power and the $COP$, (ii) there is an optimized energy barrier height in nanowires for which the cooling performance is optimized. For bulk coolers, on the other hand, the cooling power increases as one increases the energy barrier height. (iii)  The cooling performance in nanowires deteriorates compared to bulk as a single-moded nanowire transitions to a multi-moded one. While exploring the cooling performance of the Peltier cooler, we have considered a parabolic dispersion relationship and assumed optical phonon scattering to be the dominant scattering mechanism. However, it remains to be explored how the theory is modified with different elastic and inelastic scattering  mechanisms \cite{anifil} and non-parabolic dispersion relations. In particular, it remains an interesting problem to formulate a compact parameter that can be used to speculate the cooling performance based on the energy dependence of the electronic density of states,  relaxation time and electronic transport velocity.  This paper, however, sets the stage for an exploration of Peltier cooling in nanostructures. We believe that the conclusions presented here would establish a general viewpoint to understand the basics of the design of Peltier coolers and their optimization.
\appendix
\section{NEGF equations for dissipative transport}
In case of dissipative transport in nano devices, the generalized equations for non-equilibrium Green's function formalism  (NEGF) are given by \cite{dattabook,Datta_Green,LNE}:
\small
\begin{gather}
G(\overrightarrow{k_{m}},E)=[EI-H-U-E_m-\Sigma(\overrightarrow{k_{m}},E)]^{-1} \nonumber \\
\Sigma(\overrightarrow{k_{m}},E)=\Sigma_L(\overrightarrow{k_{m}},E)+\Sigma_R(\overrightarrow{k_{m}},E)+\Sigma_S(\overrightarrow{k_{m}},E) \nonumber \\
 A(\overrightarrow{k_{m}},E)=i[G(\overrightarrow{k_{m}},E)-G^{\dagger}(\overrightarrow{k_{m}},E)] \nonumber \\
 \Gamma(\overrightarrow{k_{m}},E)=[\Sigma(\overrightarrow{k_{m}},E)-\Sigma^{\dagger}(\overrightarrow{k_{m}},E)],  
 \label{eq:negf}
\end{gather}
\normalsize
where $H$ is the discretized Hamiltonian matrix constructed using the nearest neighbour tight-binding approximation in an effective mass approach, $U$ denotes the modification in the conduction band minima due to  the embedded energy barrier and $\Sigma_{L(R)}(\overrightarrow{k_{m}},E)$ and $\Sigma_S(\overrightarrow{k_{m}},E)$ describe the effect of  coupling and scattering of the electronic wavefunction  due to contacts and inelastic events (electron-phonon interaction) respectively.  $\overrightarrow{k_{m}}$ in the above set of Eqs. denotes the transverse wavevector of the $m^{th}$ sub-band.
 $A(\overrightarrow{k_{m}},E)$ is the $1-D$ spectral function  for the $m^{th}$ sub-band and $\Gamma(\overrightarrow{k_{m}},E)$ is the broadening matrix for the $m^{th}$ sub-band at  energy $E$. 
 For moderate electron-phonon interaction, it is generally assumed that the real part of $\Sigma_S=0$. Hence, 
\begin{equation}
\Sigma_S(\overrightarrow{k_{m}},E)=i\frac{ \Gamma_S(\overrightarrow{k_{m}},E)}{2}=\Sigma^{in}_S(\overrightarrow{k_{m}},E)+\Sigma^{out}_S(\overrightarrow{k_{m}},E)
\label{eq:sigma_phonon}
\end{equation}
$ \Sigma^{in}(\overrightarrow{k_{m}},E)$ and $ \Sigma^{out}(\overrightarrow{k_{m}},E)$ are the in-scattering and the  out-scattering functions which model the rate of scattering of the  electrons due to incoherence inside the device and external contacts.

\begin{multline}
  \Sigma^{in}(\overrightarrow{k_{m}},E)=\Sigma^{in}_L(\overrightarrow{k_{m}},E)+\Sigma^{in}_R(\overrightarrow{k_{m}},E) \\
  +\Sigma^{in}_S(\overrightarrow{k_{m}},E) \nonumber 
  \end{multline}
  \begin{multline}
   \Sigma^{out}(\overrightarrow{k_{m}},E)=\Sigma^{out}_L(\overrightarrow{k_{m}},E)+\Sigma^{out}_R(\overrightarrow{k_{m}},E) \\
   +\Sigma^{out}_S(\overrightarrow{k_{m}},E), 
\label{eq:sig}
\end{multline}
The in-scattering and out-scattering functions are related to the contact quasi-Fermi distribution functions  via the equations:

\begin{multline}
 \Sigma^{in}(\overrightarrow{k_{m}},E)=\underbrace{\Gamma_L(\overrightarrow{k_{m}},E)f_L(E)}_{inflow~from~left~contact}  \\+\underbrace{\Gamma_R(\overrightarrow{k_{m}},E)f_R(E)}_{inflow~from~right~contact}+\underbrace{\Sigma^{in}_S(\overrightarrow{k_{m}},E)}_{inflow~due~to~phonons}, \nonumber
 \end{multline}
 \begin{multline}
 \Sigma^{out}(\overrightarrow{k_{m}},E)=\underbrace{\Gamma_L(\overrightarrow{k_{m}},E)\Big\{ 1-f_L(E)\Big \} }_{outflow~to~left~contact} \\+\underbrace{\Gamma_R(\overrightarrow{k_{m}},E)\Big \{ 1-f_R(E)\Big \}}_{outflow~to~right~contact}+\underbrace{\Sigma^{out}_S(\overrightarrow{k_{m}},E)}_{outflow~due~to~phonons}, 
 \label{eq:sig1}
\end{multline}
where $f_{L(R)}$ represent the quasi-Fermi distribution of left(right) contact.
For local scattering mechanisms, the rate of inelastic scattering of electrons  is dependent on the electron and the hole correlation functions $(G^n$ and $G^p)$ via:

\small
\begin{multline}
\Sigma^{in}_S(\overrightarrow{k_{m}},E)=diag \Bigg\{ D_O \times \Big[ (N+1)
\underset{\overrightarrow{q_{t}}}{\sum}G^n(\overrightarrow{k_{m}}+\overrightarrow{q_{t}},E+\hslash \omega)  \\
 N \underset{\overrightarrow{q_{t}}}{\sum}G^n(\overrightarrow{k_{m}}+\overrightarrow{q_{t}},E-\hslash \omega)\Big] \Bigg\} \nonumber
 \end{multline}
 \begin{multline}
\Sigma^{out}_S(\overrightarrow{k_{m}},E)=diag \Bigg\{D_O \times \Big[ (N+1)
\underset{\overrightarrow{q_{t}}}{\sum}G^p(\overrightarrow{k_{m}}+\overrightarrow{q_{t}},E-\hslash \omega)  \\
 N \underset{\overrightarrow{q_{t}}}{\sum}G^p(\overrightarrow{k_{m}}+\overrightarrow{q_{t}},E+\hslash \omega)\Big] \Bigg\}
 \label{eq:sig_ph}
 \end{multline}
 \normalsize
 In the above set of Eqs., $N$ denotes the average phonon number given by:
 \[
N=\frac{1}{e^{\frac{\hslash\omega_o}{k_BT_L}}-1}, 
\]
 $D_O$ is related to the optical deformation potential ($D$) via the equation:
\begin{equation}
D_O=\frac{\hslash D^2F}{2\rho\omega_o a^3},
\end{equation}
and $\hslash \omega_o$ denotes the optical phonon energy, $\omega_o$ being the optical phonon radial frequency. $\{\overrightarrow{q_t}\}$ denotes the set of transverse phonon wave vectors.\\
\indent $G^n(\overrightarrow{k_{m}},E)$ and $G^p(\overrightarrow{k_{m}},E)$ are the electron and the hole correlation functions for the $m^{th}$ sub-band. The electron and the hole correlation functions are again related to the  electron in-scattering and the electron out-scattering functions via the equations:
\begin{gather}
G^n(\overrightarrow{k_{m}},E)=G(\overrightarrow{k_{m}},E)\Sigma^{in}(\overrightarrow{k_{m}},E)G^{\dagger}(\overrightarrow{k_{m}},E) \nonumber \\
G^p(\overrightarrow{k_{m}},E)=G(\overrightarrow{k_{m}},E)\Sigma^{out}(\overrightarrow{k_{m}},E)G^{\dagger}(\overrightarrow{k_{m}},E) \nonumber \\
 \label{eq:correlation} 
\end{gather}
 Solving the  dynamics of the entire system involves a self consistent solution of \eqref{eq:negf},  \eqref{eq:sig}, \eqref{eq:sig_ph} and  \eqref{eq:correlation}.
The electron density and current at the grid point $j$ can be calculated from the above equations as:
\[
n_j=\underset{m}{\sum}\int\frac{[G^n(\overrightarrow{k_{m}},E)dE]}{\pi aA}
\]
\begin{multline}
I^{j\rightarrow j+1}=\underset{k_m}{\sum}i\frac{e}{\pi \hslash} \int[
G^n_{j+1,j}(\overrightarrow{k_{m}},E)H_{j,j+1}(E)  \\ 
-H_{j+1,j}(E)G^n_{j,j+1}(\overrightarrow{k_{m}},E) ]dE,
\label{eq:currentnegf}
\end{multline}
where $a$ is the distance between two adjacent grid points and $A$ is the cross sectional area of the device. $\hslash k_m$ denotes the transverse momentum of the electrons in the $m^{th}$ sub-band. The summations in \eqref{eq:currentnegf} run over all the sub-bands available for conduction.

The heat current flowing through the device from the $j^{th}$ point to the $(j+1)^{th}$ point is given by:

\begin{multline}
I_Q^{j\rightarrow j+1}=\underset{k_m}{\sum}\frac{i}{\pi \hslash} \times  \int E[
G^n_{j+1,j}(\overrightarrow{k_{m}},E) \\ H_{j,j+1}(E)-H_{j+1,j}(E)
G^n_{j,j+1}(\overrightarrow{k_{m}},E) ]dE,  
\label{eq:heatcurrentnegf}
\end{multline}

\bibliography{apssamp}

\providecommand{\noopsort}[1]{}\providecommand{\singleletter}[1]{#1}
\begin{thebibliography}{10}

\bibitem{highfom1}
G.~Jeffrey Snyder and Eric~S. Toberer.
\newblock Complex thermoelectric materials.
\newblock {\em Nat Mater}, 7(2):105--114, Feb 2008.

\bibitem{highfom2}
Paothep Pichanusakorn and Prabhakar Bandaru.
\newblock Nanostructured thermoelectrics.
\newblock {\em Materials Science and Engineering: R: Reports}, 67(2–4):19 --
  63, 2010.

\bibitem{thermoinnano}
Arun Majumdar.
\newblock Thermoelectricity in semiconductor nanostructures.
\newblock {\em Science}, 303(5659):777--778, 2004.

\bibitem{aniket}
Aniket Singha, Subhendra~D. Mahanti, and Bhaskaran Muralidharan.
\newblock Exploring packaging strategies of nano-embedded thermoelectric
  generators.
\newblock {\em AIP Advances}, 5(10), 2015.

\bibitem{anifil}
Aniket Singha and Bhaskaran Muralidharan.
\newblock Incoherent scattering can favorably influence energy filtering in
  nanostructured thermoelectrics.
\newblock {\em Scientific Reports}, 7(1):7879, 2017.

\bibitem{snyder_thompson}
G.~Jeffrey Snyder, Eric~S. Toberer, Raghav Khanna, and Wolfgang Seifert.
\newblock Improved thermoelectric cooling based on the thomson effect.
\newblock {\em Phys. Rev. B}, 86:045202, Jul 2012.

\bibitem{cooling_ref1}
Y.~Apertet, H.~Ouerdane, A.~Michot, C.~Goupil, and Ph. Lecoeur.
\newblock On the efficiency at maximum cooling power.
\newblock {\em EPL (Europhysics Letters)}, 103(4):40001, 2013.

\bibitem{cooling_ref2}
Ali Shakouri and John~E. Bowers.
\newblock Heterostructure integrated thermionic coolers.
\newblock {\em Applied Physics Letters}, 71(9):1234--1236, 1997.

\bibitem{cooling_ref3}
Ali Shakouri Edwin Y. Lee D. L. Smith Venky Narayanamurti John~E. Bowers.
\newblock Thermoelectric effects in submicron heterostructure barriers.
\newblock {\em Microscale Thermophysical Engineering}, 2(1):37--47, 1998.

\bibitem{cooling_ref4}
Xiaofeng Fan, Gehong Zeng, E.~Croke, G.~Robinson, C.~LaBounty, A.~Shakouri, and
  J.~E. Bowers.
\newblock N- and p-type sige/si superlattice coolers.
\newblock In {\em ITHERM 2000. The Seventh Intersociety Conference on Thermal
  and Thermomechanical Phenomena in Electronic Systems (Cat. No.00CH37069)},
  volume~1, page 307, 2000.

\bibitem{cooling_ref5}
X.~Fan, G.~Zeng, E.~Croke, C.~LaBounty, D.~Vashaee, A.~Shakouri, and J.~E.
  Bowers.
\newblock High cooling power density sige/si microcoolers.
\newblock {\em Electronics Letters}, 37(2):126--127, Jan 2001.

\bibitem{cooling_ref6}
Raseong Kim, Changwook Jeong, and Mark~S. Lundstrom.
\newblock On momentum conservation and thermionic emission cooling.
\newblock {\em Journal of Applied Physics}, 107(5):054502, 2010.

\bibitem{cooling_ref7}
Jia-pei Zhu and Gao-xiang Li.
\newblock Ground-state cooling of a nanomechanical resonator with a triple
  quantum dot via quantum interference.
\newblock {\em Phys. Rev. A}, 86:053828, Nov 2012.

\bibitem{cooling_ref8}
Zeng-Zhao Li, Shi-Hua Ouyang, Chi-Hang Lam, and J.~Q. You.
\newblock Cooling a nanomechanical resonator by a triple quantum dot.
\newblock {\em EPL (Europhysics Letters)}, 95(4):40003, 2011.

\bibitem{cooling_ref9}
Francesco Giazotto, Tero~T. Heikkil\"a, Arttu Luukanen, Alexander~M. Savin, and
  Jukka~P. Pekola.
\newblock Opportunities for mesoscopics in thermometry and refrigeration:
  Physics and applications.
\newblock {\em Rev. Mod. Phys.}, 78:217--274, Mar 2006.

\bibitem{cooling_ref10}
H.~L. Edwards, Q.~Niu, and A.~L. de~Lozanne.
\newblock A quantum‐dot refrigerator.
\newblock {\em Applied Physics Letters}, 63(13):1815--1817, 1993.

\bibitem{cooling_ref11}
K.~A. Chao, Magnus Larsson, and A.~G. Mal’shukov.
\newblock Room-temperature semiconductor heterostructure refrigeration.
\newblock {\em Applied Physics Letters}, 87(2):022103, 2005.

\bibitem{whitney2}
Robert~S. Whitney.
\newblock Finding the quantum thermoelectric with maximal efficiency and
  minimal entropy production at given power output.
\newblock {\em Phys. Rev. B}, 91:115425, Mar 2015.

\bibitem{whitney}
R.~S. {Whitney}.
\newblock {Most Efficient Quantum Thermoelectric at Finite Power Output}.
\newblock {\em Physical Review Letters}, 112(13):130601, April 2014.

\bibitem{phonon1}
N.~Mingo and D.~A. Broido.
\newblock Lattice thermal conductivity crossovers in semiconductor nanowires.
\newblock {\em Phys. Rev. Lett.}, 93:246106, Dec 2004.

\bibitem{phonon2}
N.~Mingo.
\newblock Thermoelectric figure of merit and maximum power factor in iii–v
  semiconductor nanowires.
\newblock {\em Applied Physics Letters}, 84(14):2652--2654, 2004.

\bibitem{phonon3}
Feng Zhou, Jeannine Szczech, Michael~T. Pettes, Arden~L. Moore, Song Jin, and
  Li~Shi.
\newblock Determination of transport properties in chromium disilicide
  nanowires via combined thermoelectric and structural characterizations.
\newblock {\em Nano Letters}, 7(6):1649--1654, 2007.
\newblock PMID: 17508772.

\bibitem{phonon4}
Feng Zhou, Arden~L Moore, Michael~T Pettes, Yong Lee, Jae~Hun Seol, Qi~Laura
  Ye, Lew Rabenberg, and Li~Shi.
\newblock Effect of growth base pressure on the thermoelectric properties of
  indium antimonide nanowires.
\newblock {\em Journal of Physics D: Applied Physics}, 43(2):025406, 2010.

\bibitem{phonon5}
Akram~I Boukai, Yuri Bunimovich, Jamil Tahir-Kheli, Jen-Kane Yu, William~A.l
  Goddard, and James~R Heath.
\newblock Silicon nanowires as efficient thermoelectric materials.
\newblock {\em Nature}, 451:168--171, 2008.

\bibitem{phonon6}
Allon~I. Hochbaum, Renkun Chen, Raul~Diaz Delgado, Wenjie Liang, Erik~C
  Garnett, Mark Najarian, Arun Majumdar, and Peidong Yang.
\newblock Enhanced thermoelectric performance of rough silicon nanowires.
\newblock {\em Nature Publishing Group}, 451:163--167, 2008.

\bibitem{phonon7}
A.~Balandin, A.~Khitun, J.L. Liu, K.L. Wang, T.~Borca-Tasciuc, and G.~Chen.
\newblock Optimization of the thermoelectric properties of low-dimensional
  structures via phonon engineering.
\newblock In {\em Eighteenth International Conference on Thermoelectrics},
  pages 189--192, Aug 1999.

\bibitem{superlattice1}
G.~Chen.
\newblock Thermal conductivity and ballistic-phonon transport in the
  cross-plane direction of superlattices.
\newblock {\em Phys. Rev. B}, 57:14958--14973, Jun 1998.

\bibitem{superlattice2}
T.~Koga, S.~B. Cronin, M.~S. Dresselhaus, J.~L. Liu, and K.~L. Wang.
\newblock Experimental proof-of-principle investigation of enhanced z3dt in
  (001) oriented si/ge superlattices.
\newblock {\em Applied Physics Letters}, 77(10), 2000.

\bibitem{nanoflake_heat}
Bruce~L. Davis and Mahmoud~I. Hussein.
\newblock Nanophononic metamaterial: Thermal conductivity reduction by local
  resonance.
\newblock {\em Phys. Rev. Lett.}, 112:055505, Feb 2014.

\bibitem{nanowire_heat1}
Ying Pan, Guo Hong, Shyamprasad~N. Raja, Severin Zimmermann, Manish~K. Tiwari,
  and Dimos Poulikakos.
\newblock Significant thermal conductivity reduction of silicon nanowire
  forests through discrete surface doping of germanium.
\newblock {\em Applied Physics Letters}, 106(9):093102, 2015.

\bibitem{nanowire_heat2}
Joseph~P. Feser, Jyothi~S. Sadhu, Bruno~P. Azeredo, Keng~H. Hsu, Jun Ma,
  Junhwan Kim, Myunghoon Seong, Nicholas~X. Fang, Xiuling Li, Placid~M.
  Ferreira, Sanjiv Sinha, and David~G. Cahill.
\newblock Thermal conductivity of silicon nanowire arrays with controlled
  roughness.
\newblock {\em Journal of Applied Physics}, 112(11):114306, 2012.

\bibitem{ieeecool}
C.~J. Mole, D.~V. Foster, and R.~A. Feranchak.
\newblock Thermoelectric cooling technology.
\newblock {\em IEEE Transactions on Industry Applications}, IA-8(2):108--125,
  March 1972.

\bibitem{dattabook}
Supriyo Datta.
\newblock {\em Quantum Transport:Atom to Transistor}.
\newblock Cambridge Press, 2005.

\bibitem{Datta_Green}
Supriyo Datta.
\newblock {\em {Electronic Transport in Mesoscopic Systems}}.
\newblock Cambridge University Press, May 1997.

\bibitem{LNE}
Supriyo Datta.
\newblock {\em {Lessons from nanoelectronics: a new perspective on transport}}.
\newblock Lessons from nanosciences: A lecture note series. World Scientific,
  Singapore, 2012.

\bibitem{book1}
Luca~Selmi David~Esseni, Pierpaolo~Palestri.
\newblock {\em Nanoscale MOS Transistors Semi-Classical Transport and
  Applications}.
\newblock Cambridge, 2011.

\end{thebibliography}
\bibliographystyle{unsrt}

\end{document}